\documentclass[twoside,english,3p]{elsarticle}
\usepackage[T1]{fontenc}
\usepackage[latin9]{inputenc}
\usepackage{array}
\usepackage{float}
\usepackage{graphicx}
\usepackage{setspace}

\makeatletter

\newcommand{\noun}[1]{\textsc{#1}}
\providecommand{\tabularnewline}{\\}
\floatstyle{ruled}
\newfloat{algorithm}{tbp}{loa}
\providecommand{\algorithmname}{Algorithm}
\floatname{algorithm}{\protect\algorithmname}

\makeatletter
\def\ps@pprintTitle{%
 \let\@oddhead\@empty
 \let\@evenhead\@empty
 \def\@oddfoot{}%
 \let\@evenfoot\@oddfoot}
\makeatother

\makeatother

\usepackage{babel}
\usepackage{listings}

\begin{document}

\begin{frontmatter}{}

\title{Application of Data Science to Discover Violence-Related Issues in
Iraq}

\author[focal]{Merari González}

\author[focal]{Germán H. Alférez\corref{cor1}}

\ead{harveyalferez@um.edu.mx}

\cortext[cor1]{Corresponding author}

\address[focal]{Facultad de Ingeniería y Tecnología, Universidad de Montemorelos,
Apartado 16-5, 67500 Montemorelos, N.L., Mexico}
\begin{abstract}
Data science has been satisfactorily used to discover social issues
in several parts of the world. However, there is a lack of governmental
open data to discover those issues in countries such as Iraq. This
situation arises the following questions: how to apply data science
principles to discover social issues despite the lack of open data
in Iraq? How to use the available data to make predictions in places
without data? Our contribution is the application of data science
to open non-governmental big data from the Global Database of Events,
Language, and Tone (GDELT) to discover particular violence-related
social issues in Iraq. Specifically we applied the K-Nearest Neighbors,
Näive Bayes, Decision Trees, and Logistic Regression classification
algorithms to discover the following issues: refugees, humanitarian
aid, violent protests, fights with artillery and tanks, and mass killings.
The best results were obtained with the Decision Trees algorithm to
discover areas with refugee crises and artillery fights. The accuracy
for these two events is 0.7629. The precision to discover the locations
of refugee crises is 0.76, the recall is 0.76, and the F1-score is
0.76. Also, our approach discovers the locations of artillery fights
with a precision of 0.74, a recall of 0.75, and a F1-score of 0.75.
\end{abstract}

\end{frontmatter}{}

\section{Introduction}

Data science is the generalizable extraction of knowledge from data
\citep{Dhar2013}. Data is the fuel of data science to discover the
places in which help could be provided (for instance, humanitarian
aid). However, most data from the Middle East is blocked because of
legal or technical restrictions. Specifically, Iraq is one of the
countries that is not in the regional ranking regarding to open data
\citep{OpenDataBarometer}.

In previous work, thanks to machine learning, a sub-area of artificial
intelligence and a key component of data science, it has been possible
to discover certain social behaviors or even to predict future events.
For instance, in \citep{Yonamine2012} a model was proposed to predict
certain levels of violence in districts of Afghanistan. In \citep{Bi2015},
data science was used to estimate the degree of activity and influence
of the USA, China, and USA with respect to politics, economy, trade,
culture, the military, among other areas. However, despite the effervescence
in the use of machine learning to discover interesting social trends,
the lack of data in Iraq makes it difficult to apply this approach
in that country.

Our contribution in this research work is the application of data
science to open non-governmental big data to discover particular social
issues related to violence in Iraq. We are interested in this country
because violence, refugee crises, and massacres are constant in its
territory \citep{ONU}. Specifically we applied the K-Nearest Neighbors
(KNN), Näive Bayes, Decision Trees, and Logistic Regression classification
algorithms to discover violence-related social issues in the territory
of Iraq in terms of refugees, humanitarian aid, violent protests,
fights with artillery and tanks, and mass killings. The models were
trained with open data from the Global Database of Events, Language,
and Tone (GDELT), a project sponsored by Google that contains more
than 200 million geolocated events with worldwide coverage regarding
news and important events from 1979 \citep{Leetaru2013a}. GDELT obtains
the data from news agencies such as Lexis Nexis, the Agence France-Presse,
Reuters, Associated Press, and Xinhua. It also uses the Conflict and
Mediation Event Observations (CAMEO) code, which is a content code
for electronic news \citep{Yonamine2012}.

In our experiments, the best results were obtained with the Decision
Trees algorithm to discover areas with refugee crises and artillery
fights. The accuracy value for these two events was 0.7629. The precision,
recall, and F1-score values to discover the locations of refugees
was 0.76. Also, our approach discovers the locations of artillery
fights with a precision of 0.74, a recall of 0.75, and a F1-score
of 0.75.

This paper is organized as follows. The second section presents the
related work. The third section presents the methodology. The fourth
section presents the evaluation results. The fifth section describes
the software for classification. The last section presents the conclusions
and future work.

\section{Related work}

Bi, Gao, Wang y Cao \citep{Bi2015} used GDELT data to estimate the
degree of activity and influence of countries regarding politics,
economy, commerce, culture, militia, among others. In their research
work, countries such as the United States, China, and Russia stand
out. Also Kumar, Benigni. and Carley \citep{Kumar2016} used data
from GDETL to analyze the perception of the population towards cyber
attacks in the United States through graphs. By observing them, they
concluded that cyber attacks have decreased in relation to changes
in the country's cyber policy. In another research work, Su, Lan,
Lin, Comfort, and Joshi \citep{Su2016} describe the data analysis
of the earthquake response in Nepal in 2015. The data were obtained
from GDELT and the results show the increase and decrease of people's
interest in contributing through donations. They also explain the
support of the government and the effectiveness of international and
local organizations. They also describe the importance of careful
monitoring and prompt attention after disasters. As far as we know,
there is not previous related work about using data science applied
to GDELT data to discover social issues related to violence of people
in Iraq.

\section{Methodology}

Data science follows a general process that includes data collection,
data cleaning, data analysis, and modeling \citep{Phethean2016}.
In data science, the data obtained is processed differently than in
traditional approaches, so a new methodology must be applied to extract
the necessary knowledge. In fact, the use of a data-oriented methodology
allows analysts to develop and evaluate predictive models in an efficient
way \citep{Murphree2016}. To this end, we used the IBM Foundational
Methodology for Data Science, composed of ten stages that represent
an iterative process \citep{Rollins2015}. The first seven stages
are described in this section in the context of our research work.
The evaluation, deployment, and feedback stages are described in the
next sections.

\subsection{Problem understanding}

According to the Open Data Barometer \citep{OpenDataBarometer}, there
are few countries in the Middle East and North Africa region with
open data initiatives. This is due to the low participation of civilians
in this kind of initiatives and little pressure of governments to
make their data public. In this research work, Iraq was taken as a
case study because that is one of the countries in the Middle East
with unavailable open data due to legal restrictions.

\subsection{Analytic approach}

Machine learning was chosen as the mechanism to understand and analyze
data regarding refugees, humanitarian aid, violent protests, fights
with artillery and tanks, and mass killings in Iraq. These events
were selected due to their constant mention in the news from that
country \citep{ONU}. In the experiments, Python was used together
with Scikit Learn\footnote{https://scikit-learn.org/stable/} to program
the following machine learning algorithms: KNN, Decision Trees, Näive
Bayes, and Logistic Regression.

\subsection{Data requirements}

In this step, a query was executed in Big Query to obtain the GDELT
data about the events studied in Iraq. The 42,027 records retrieved
during the query include the latitude and longitude in Iraq where
the news happened. The query covered data from 2012 to 2015. Specifically,
Table 1 shows the description of the variables used in Big Query,
where the \noun{actor} refers to the entity involved in the event.
The codes used to get news from particular latitudes and longitudes
are also shown.

\begin{table}[h]
\caption{Description of variables used in the query \citep{GDELT2013,Schrodt2012}.}

\centering{}%
\begin{tabular}{|>{\centering}p{4cm}|>{\centering}p{4cm}|>{\centering}p{2cm}|}
\hline 
\textbf{Code} & \textbf{Description} & \textbf{Type}\tabularnewline
\hline 
\hline 
ActionGeo\_CountryCode & Country where the event happened and the event & String\tabularnewline
\hline 
Actor1Geo\_Lat & Latitude of the location of the actor & Float\tabularnewline
\hline 
Actor1Geo\_Long & Longitude of the actor & Float\tabularnewline
\hline 
Actor1Type1Code & Type or role of the actor (e.g. refugees) & String\tabularnewline
\hline 
EventCode & Entities related to the event & String\tabularnewline
\hline 
Year & Year of the event & Integer\tabularnewline
\hline 
\end{tabular}
\end{table}

\subsection{Data collection}

The results obtained with Big Query were downloaded in the comma separated
values (CSV) format. Google Maps was used to get the data from Iraq.
The geolocation points that were considered to get the data were the
following: latitude greater than 29.12 and lower than 37.29 and longitude
greater than 39.22 and lower than 48.48. The \emph{ActionGeo\_CountryCode}
corresponds to IZ (Iraq in CAMEO code). Figure 1 shows the limits
in terms of extreme latitudes and longitudes of the territory of Iraq.
These points were obtained with Google Maps.

\begin{figure}[h]
\begin{centering}
\includegraphics[width=0.8\columnwidth]{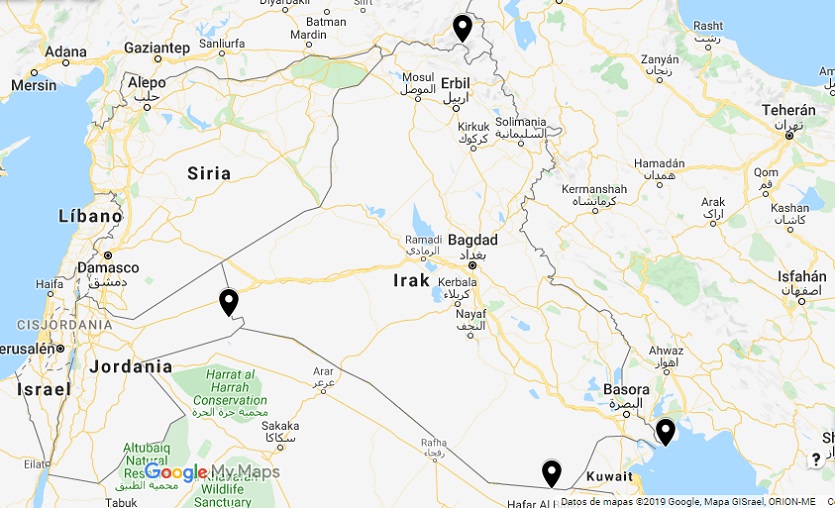}
\par\end{centering}
\caption{Extreme latitude and longitude points for Iraq.}
\end{figure}

\begin{singlespace}
A query example in Big Query to get data from Iraq from 2012 to 2015
according to the aforementioned latitudes and longitudes is shown
in Listing 1.
\end{singlespace}

\begin{algorithm}[H]
\begin{lstlisting}
SELECT Actor1Type1Code,Year, 
ActionGeo_CountryCode, Actor1Geo_Lat,
Actor1Geo_Long, EventCode

FROM [gdelt-bq:full.events]

WHERE Actor1Type1Code="REF" 

AND (Year> 2011 AND Year < 2016) 
AND (Actor1Geo_Lat > 29.12 
AND Actor1Geo_Lat < 37.29) 
AND (Actor1Geo_Long > 39.22 
AND Actor1Geo_Long < 48.48) 
AND Actor1Geo_Lat IS NOT NULL 
AND Actor1Geo_Long IS NOT NULL 
\end{lstlisting}

\caption{Query example in Big Query to get the data related to refugees in
Iraq from GDELT.}
\end{algorithm}

Table 2 shows the distribution of the 42,027 records obtained from
the query. The CSV file with the results of the query is available
online\footnote{https://drive.google.com/drive/folders/0B6LEG8jNfAY9MEJBTkZoYlRIZHc}.

\begin{table}[H]
\caption{GDELT data obtained with BigQuery.}

\centering{}%
\begin{tabular}{|c|c|c|}
\hline 
\textbf{Event} & \textbf{Description} & \textbf{Period 2012-2015}\tabularnewline
\hline 
\hline 
073 & Provide humanitarian aid & 10,414\tabularnewline
\hline 
145 & Violent protests & 3,068\tabularnewline
\hline 
194 & Fight with artillery and tanks & 13,247\tabularnewline
\hline 
202 & Engage in mass killings & 1,822\tabularnewline
\hline 
REF & Refugees & 13,476\tabularnewline
\hline 
\multicolumn{2}{|c|}{\textbf{Total number of records}} & \textbf{42,027}\tabularnewline
\hline 
\end{tabular}
\end{table}

\subsection{Data understanding}

We used Google Maps to understand the data obtained from GDELT. The
results of these events are shown in Figure 2 to Figure 6\footnote{https://www.google.com/maps/d/edit?hl=es\&hl=es\&authuser\\=0\&authuser=0\&mid=1uEeGUsxz38AdaYyYeHCJdWt5m\\G8\&ll=33.137338087845784\%2C43.81860000000006\&z=6}.
Some points in the maps exceed the limits established for latitude
and longitude in Iraq. This is because these distances were obtained
manually using the extreme limits on the map, as described in the
previous step (see Figure 1). In the maps, the violence-related events
studied in this research work have a greater incidence in the north
and east of the country. It is also interesting to see that news related
to humanitarian aid and artillery fights cover most of the country.

Although the simple visualization of the data can be used to find
patterns, it is important to notice that there are areas where data
is not available according to the events studied (i.e., blank areas).
It is where machine learning comes into play to discover violence-related
issues in areas with insufficient data.

\begin{figure}
\begin{centering}
\includegraphics[scale=0.4]{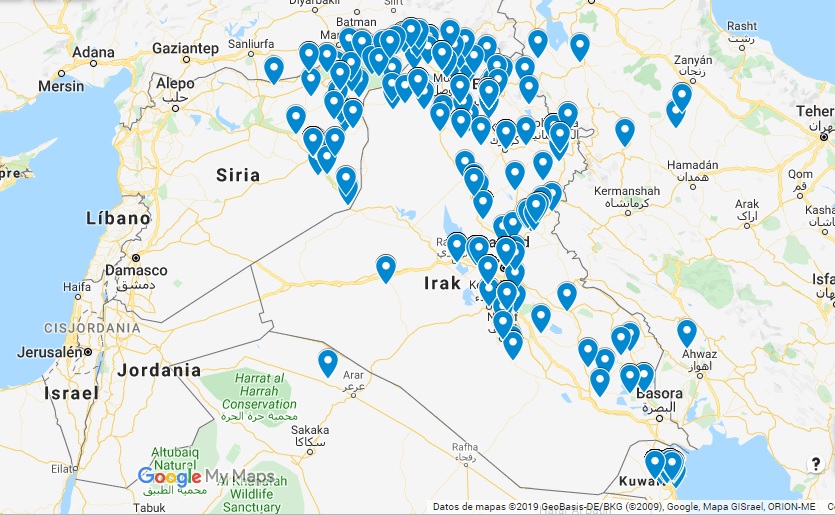}
\par\end{centering}
\caption{Map of news related to refugees.}
\end{figure}

\begin{figure}
\begin{centering}
\includegraphics[scale=0.4]{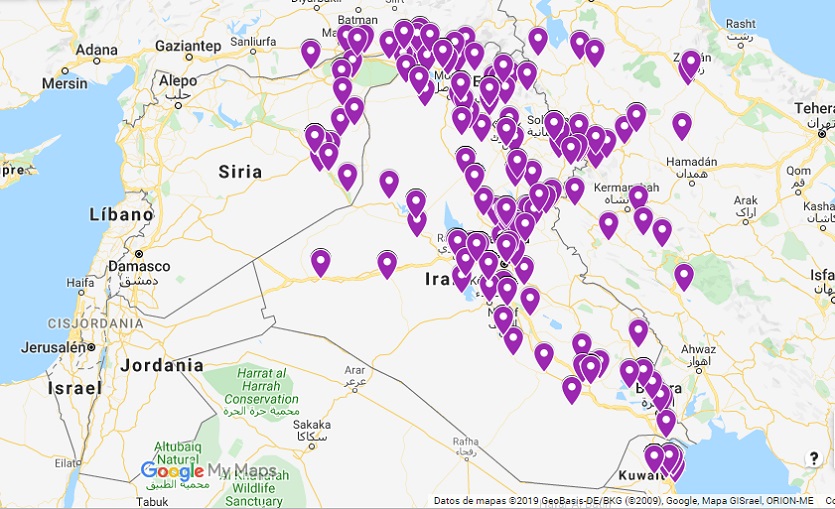}
\par\end{centering}
\caption{Map of news related to humanitarian aid.}
\end{figure}

\begin{figure}
\begin{centering}
\includegraphics[scale=0.4]{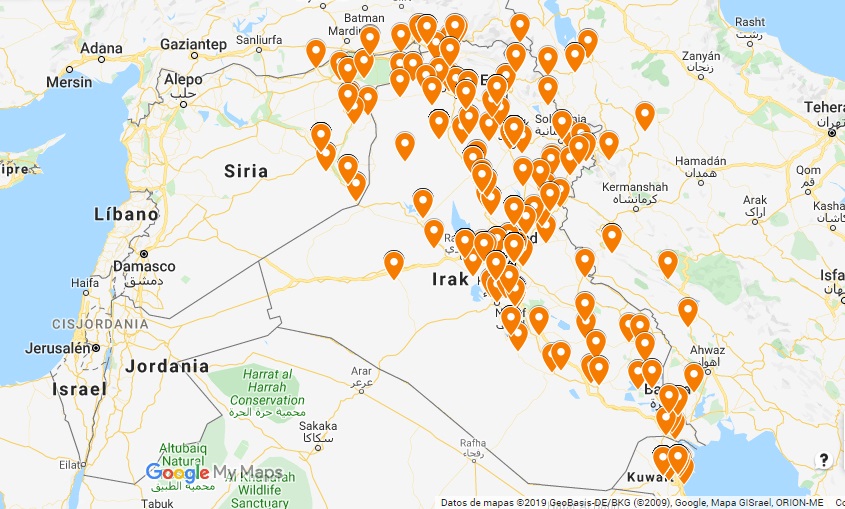}
\par\end{centering}
\caption{Map of news related to violent protests.}
\end{figure}

\begin{figure}[H]
\begin{centering}
\includegraphics[scale=0.4]{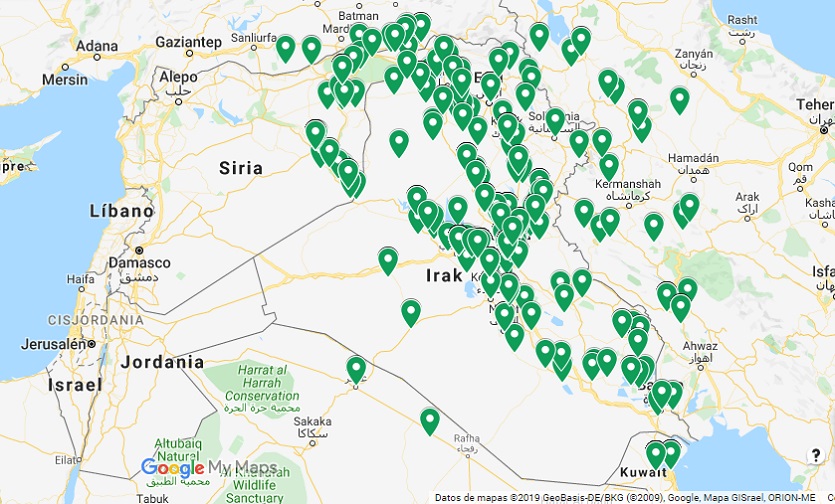}
\par\end{centering}
\caption{Map of news related to fights with artillery and tanks.}
\end{figure}

\begin{figure}[H]
\begin{centering}
\includegraphics[scale=0.4]{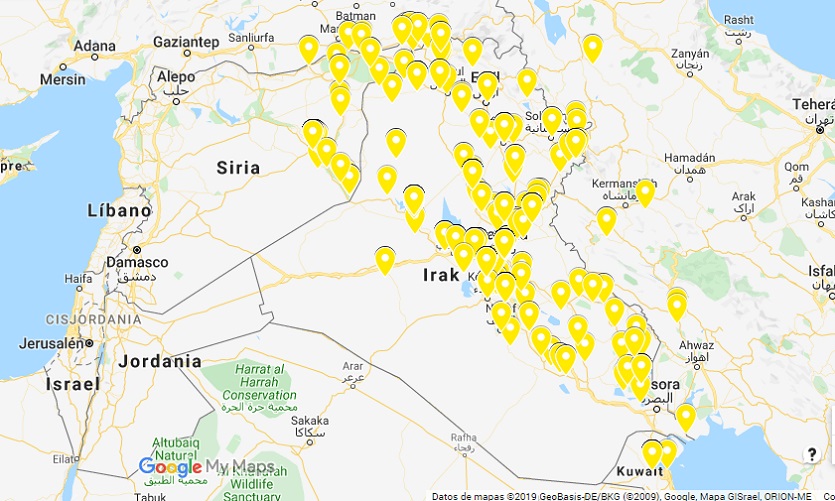}
\par\end{centering}
\caption{Map of news related to mass killings.}
\end{figure}

\subsection{Data preparation and Modeling}

The String GDELT code values were converted to Integer values to run
the experiments (see Table 1). It was a necessary step because machine
learning algorithms work with numerical data. Specifically, the \noun{ActionGeo\_CountryCode}
code value for Iraq had a String data type that was converted to Integer.
Moreover, the values for the \noun{Actor1Type1Code} code had the String
data type as returned by GDELT. The values for this variable were
also converted to Integer values (for instance, the \noun{REF} actor
type code was converted to 0). Also, the following Integer values
were assigned to the \noun{EventCode} code: for the refugees event,
the value is 0; for the humanitarian aid event, the value is 073;
for the violent protests event, the value is 145; for the fights with
artillery and tanks event, the value is 194; and for the engage in
mass killings event, the value is 202.

In the modeling step, predictive models were generated with machine
learning, using the following classification algorithms: KNN, Näive
Bayes, Decision Trees, and Logistic Regression. These algorithms were
chosen because they are popular in supervised learning \citep{Wu2008}
and are also insensitive to outliers \citep{Micenkova2014,Barabas2009,Harrington2012,Iconiq2016}.
To train the models, the studied events were used as classes and the
latitude and longitude values were considered as features.

\section{Evaluation}

The evaluation was performed on a Lenovo laptop with the following
characteristics: AMD A8-7410 APU processor with AMD Radeon R5 Graphics,
8 GB RAM, and Windows 10, 64-bits operating system. Also, Python version
2.7.12 was used along with Anaconda version 4.2.0, which contains
Pandas and Numpy. The dataset was split into two groups: 70\% of the
instances or records were used for training and the remaining 30\%
for evaluating the model.

The experiments consisted in applying the KNN, Decision Trees, Näive
Bayes, and Logistic Regression algorithms to the dataset obtained
from GDELT. We calculated all the possible combinations of available
events for each experiment based on the following equation:

\[
C(n,r)=^{n}C_{r}=nCr=(nr)=n!/r!(n-r)!
\]

In this equation, \emph{n} stands for the number of entities that
can be chosen in a particular experiment\emph{ }and \emph{r} is the
way in which these events can be chosen. 26 combinations of events
were obtained, which is equivalent to 26 datasets that combine the
4 events studied (i.e., the classes). With these datasets, training
was carried out with each algorithm, reaching a total of 104 tests
(i.e., 4 classification algorithms multiplied by 26 datasets). The
files that include the dataset of the experiments are available online\footnote{https://drive.google.com/drive/folders/0B6LEG8jNfAY9ME\\JBTkZoYlRIZHc}.
Table 3 shows the combinations obtained from the events and the number
of instances per combination.

\begin{table}
\caption{Number of instances per dataset with different combinations of events.}

\centering{}%
\begin{tabular}{|c|c|}
\hline 
\textbf{Events} & \textbf{Number of Instances}\tabularnewline
\hline 
\hline 
\multicolumn{2}{|c|}{\emph{Two events}}\tabularnewline
\hline 
73, 145 & 13,482\tabularnewline
\hline 
73, 194 & 23,661\tabularnewline
\hline 
73, 202 & 12,236\tabularnewline
\hline 
0,73 & 23,890\tabularnewline
\hline 
145, 194 & 16,315\tabularnewline
\hline 
145, 202 & 4,890\tabularnewline
\hline 
0, 145 & 16,315\tabularnewline
\hline 
194, 202 & 15,069\tabularnewline
\hline 
0, 194 & 26,723\tabularnewline
\hline 
0, 202 & 15,298\tabularnewline
\hline 
\multicolumn{2}{|c|}{\emph{Three events}}\tabularnewline
\hline 
73, 145, 194 & 26,729\tabularnewline
\hline 
73, 145, 202 & 15,304\tabularnewline
\hline 
0, 73, 145 & 26,958\tabularnewline
\hline 
73, 194, 202 & 25,483\tabularnewline
\hline 
0, 73, 194 & 37,137\tabularnewline
\hline 
0, 73, 202 & 25,712\tabularnewline
\hline 
145, 194, 202 & 18,138\tabularnewline
\hline 
0, 145, 194 & 29,791\tabularnewline
\hline 
0, 145, 202 & 18,366\tabularnewline
\hline 
0, 194, 202 & 28,545\tabularnewline
\hline 
\multicolumn{2}{|c|}{\emph{Four events}}\tabularnewline
\hline 
73, 145, 194, 202 & 28,551\tabularnewline
\hline 
0, 73, 145, 194 & 40,205\tabularnewline
\hline 
0, 73, 145, 202 & 28,780\tabularnewline
\hline 
0, 73, 194, 202 & 38,959\tabularnewline
\hline 
0, 145, 194, 202 & 31,613\tabularnewline
\hline 
\end{tabular}
\end{table}

The results obtained in the experiments were evaluated in terms of
accuracy, precision, recall, and F1-score. With respect to accuracy,
the expected value to approve the classification model had to be as
close as possible to the value of 1 in order to obtain the highest
accuracy in the classification \citep{Scikit-Learn}. For precision
and recall values, values close to 1 were also considered appropriate
to obtain a correct classification. Specifically, a low value in precision
indicates a high number of false positives \citep{Brownlee2016}.
A low recall value indicates a high number of false negatives \citep{Hackeling2014}.

The experiments carried out with the refugee and fight with artillery
and tanks events obtained the best results in terms of accuracy, precision,
recall and F1-score. The results of the experiments with two events
that obtained the \emph{best scores} are described in the following
subsections. The source code of the programs that were used in the
experiments is available online\footnote{https://drive.google.com/drive/u/0/folders/0B6LEG8jNfAY9\\cUNtNm9mVkV4WFU}.

\subsection{Experiment 1 - KNN algorithm}

When using KNN, the best result was obtained with two events (or classes)
for the dataset of refugees (event 0 with 4,105 instances) and artillery
fight (event 194 with 3,912 instances) with a total of 8,017 instances.
Table 4 describes the average of the results with respect to precision,
recall, and F1-score. In this experiment, an accuracy of 0.7545 and
the results for the precision, recall, and F1-score measurements were
appropriate in this study.

\begin{table}[H]
\caption{Results of the KNN algorithm - refugees (event 0) and fights with
artillery and tanks (event 194).}

\centering{}%
\begin{tabular}{|c|c|c|c|}
\hline 
\textbf{Event} & \textbf{Precision} & \textbf{Recall} & \textbf{F1-score}\tabularnewline
\hline 
\hline 
0 & 0.75 & 0.75 & 0.75\tabularnewline
\hline 
194 & 0.74 & 0.74 & 0.74\tabularnewline
\hline 
\end{tabular}
\end{table}

\subsection{Experiment 2 - Decision Trees algorithm}

In the second experiment, the Decision Trees algorithm was applied.
The best result corresponded to the classification of the refugee
event (event 0, with 4,105 instances) and the fight with artillery
and tanks event (event 194, with 3,912 instances). An accuracy of
0.7629 was obtained and the precision, recall, and F1-score values
were between 0.74 and 0.76, as shown in Table 5. There is a similarity
of these results compared to the ones in the experiments with the
KNN algorithm in Table 4.

\begin{table}[H]
\caption{Results of the Decision Tree algorithms - refugees (event 0) and the
fights with artillery and tanks (event 194).}

\centering{}%
\begin{tabular}{|c|c|c|c|}
\hline 
\textbf{Event} & \textbf{Precision} & \textbf{Recall} & \textbf{F1-score}\tabularnewline
\hline 
\hline 
0 & 0.76 & 0.76 & 0.76\tabularnewline
\hline 
194 & 0.74 & 0.75 & 0.75\tabularnewline
\hline 
\end{tabular}
\end{table}

\subsection{Experiment 3 - Näive Bayes algorithm}

In the third experiment, the Näive Bayes algorithm was used. The best
result was obtained by applying this classifier to the provide humanitarian
aid event (event 73 with 3,124 instances) and the engage in mass killings
event (event 202 with 513 instances). For these events, the accuracy
value was 0.8510. However, the values for precision, recall, and F1-score
in the engage in mass killings event was 0, as shown in Table 6.

\begin{table}[H]
\caption{Results of Näive Bayes algorithm - provide humanitarian aid (event
73) and engage in mass killings (event 202).}

\centering{}%
\begin{tabular}{|c|c|c|c|}
\hline 
\textbf{Event} & \textbf{Precision} & \textbf{Recall} & \textbf{F1-score}\tabularnewline
\hline 
\hline 
73 & 0.85 & 1.00 & 0.92\tabularnewline
\hline 
202 & 0.00 & 0.00 & 0.00\tabularnewline
\hline 
\end{tabular}
\end{table}

The Näive Bayes algorithm was also used to create a model for the
refugee event (event 0 with 4,116 instances) and the violent protests
event (event 145 with 915 instances). An accuracy value of 0.8145
was obtained. However, the accuracy in the case of the protest violently
event was 0, as shown in Table 7. This demonstrates that a high value
in accuracy is not enough for the selection of the algorithm as stated
in related work \citep{Brownlee2014}.

\begin{table}[H]
\caption{Results of the Näive Bayes algorithm - refugees (event 0) and violent
protests (event 145).}

\centering{}%
\begin{tabular}{|c|c|c|c|}
\hline 
\textbf{Event} & \textbf{Precision} & \textbf{Recall} & \textbf{F-score}\tabularnewline
\hline 
\hline 
0 & 0.82 & 1.00 & 0.90\tabularnewline
\hline 
145 & 0.00 & 0.00 & 0.00\tabularnewline
\hline 
\end{tabular}
\end{table}

\subsection{Experiment 4 - Logistic Regression algorithm}

In this experiment, the Logistic Regression algorithm was applied
to the fight with artillery and tanks event (event 194 with 4,022
instances) and engage in mass killings event (event 202 with 499 instances).
The generated model obtained an accuracy value of 0.8896. However,
the engage in mass killings event obtained low values in terms of
precision, recall, and F1-score as shown in Table 8.

\begin{table}[H]
\caption{Results of the Logistic Regression algorithm - fight with artillery
and tanks (event 194) and engage in mass killings (event 202).}

\centering{}%
\begin{tabular}{|c|c|c|c|}
\hline 
\textbf{Event} & \textbf{Precision} & \textbf{Recall} & \textbf{F1-score}\tabularnewline
\hline 
\hline 
194 & 0.89 & 1.00 & 0.94\tabularnewline
\hline 
202 & 0.00 & 0.00 & 0.00\tabularnewline
\hline 
\end{tabular}
\end{table}

\subsection{Discussion}

According to the experiments presented in the last sections, the Decision
Trees algorithm stands out for the classification of the refugee event
(event 0) and fight with artillery and tanks event (event 194) - see
Section 4.2. By applying this algorithm, an accuracy of 0.7629 was
obtained. The refugee event is classified with a precision value of
0.76, a recall value of 0.76, and a F1-score value of 0.76. The fight
with artillery and tanks event (event 194) was classified with a precision
value of 0.74, a recall value of 0.75, and a F1-score value of 0.75.
Due to these results, the predictive model created with the Decision
Trees was selected to carried out the classifications of violence-related
social issues in Iraq.

To corroborate the validity of the model created with the Decision
Trees algorithm, in addition to the cross-validation evaluation presented
in the previous section, the results of the classification model were
compared with the geolocation results obtained from maps. First, we
compared the results of the model with the High Commissioner of Nations
United (UNHCR) map for refugees in Iraq (see the blue areas in Figure
7). Second, we compared the results of the model with the Liveumap
(Live Universal Awareness Map) in terms of zones of fights with artillery
and tanks (see Figure 8). The icons in this map indicate the places
of the most outstanding news.

\begin{figure}
\begin{centering}
\includegraphics[scale=0.7]{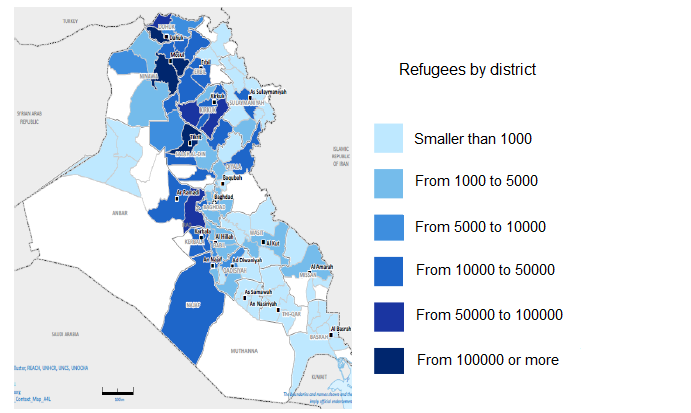}
\par\end{centering}
\caption{UNHCR map of refugees in Iraq \citep{ONU}.}
\end{figure}

\begin{figure}[H]
\begin{centering}
\includegraphics[scale=0.55]{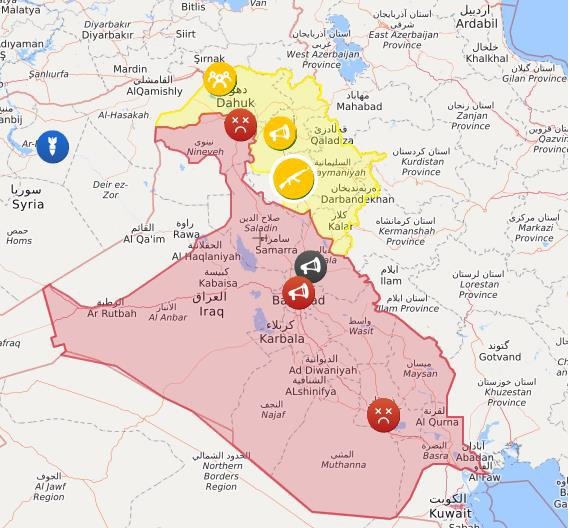}
\par\end{centering}
\caption{Liveumap map of fight with artillery and tanks in Iraq \citep{Liveuamap}.}
\end{figure}

Figure 9 shows the classification results of the areas with the highest
number of refugees by district in Figure 7 (the zones with darker
zones). These points correspond to: , Hilla, Suleimaniya, Saladin,
Tal Afar, Ninaua, Kerbala, and Erbil. In the experiments, the software
correctly classifies the refugees areas compared to the official map
of the UNHCR. Figure 10 shows the classification results of the areas
with points related to fights with artillery and tanks based on Figure
8. These points correspond to the following cities: Ramadi, Bagdad,
Kirkuk, Al-Hawija, Border Al-Kaim, Ambar, and Rawa. The software correctly
classifies this event in these locations.

\begin{figure}[H]
\begin{centering}
\begin{tabular}{cc}
a) \includegraphics[scale=0.5]{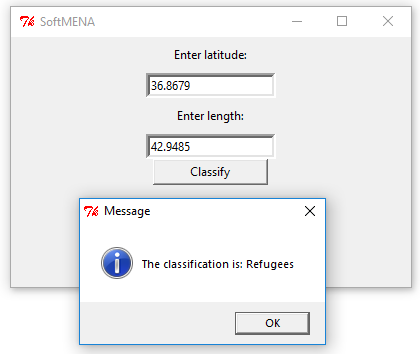} & b) \includegraphics[scale=0.5]{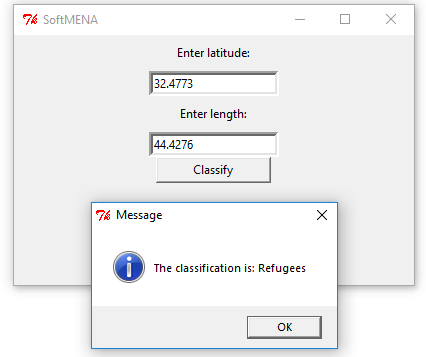}\tabularnewline
c) \includegraphics[scale=0.5]{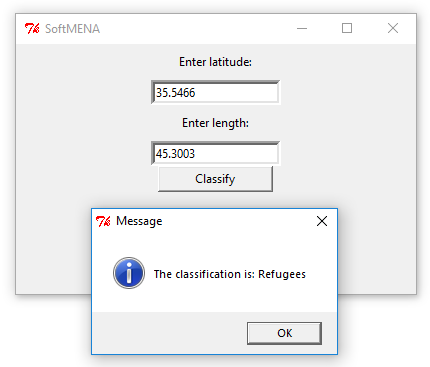} & d) \includegraphics[scale=0.5]{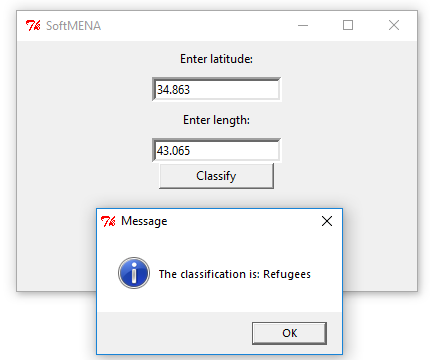}\tabularnewline
e)\includegraphics[scale=0.5]{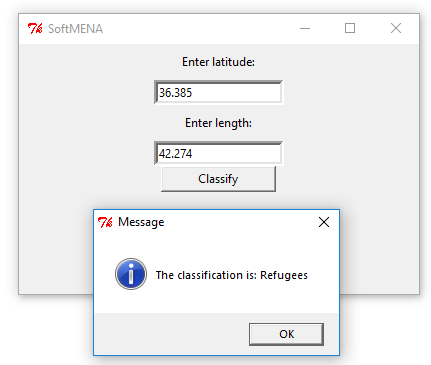} & f) \includegraphics[scale=0.5]{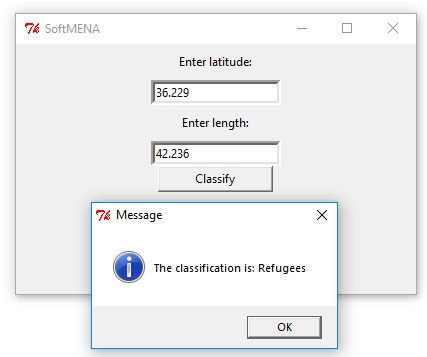}\tabularnewline
g) \includegraphics[scale=0.5]{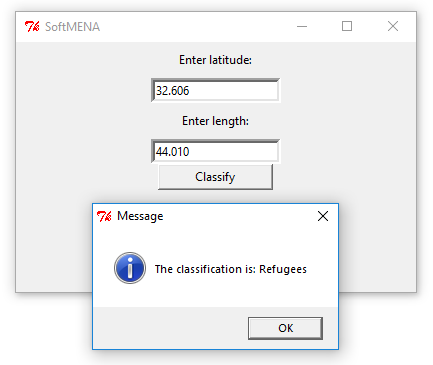} & h) \includegraphics[scale=0.5]{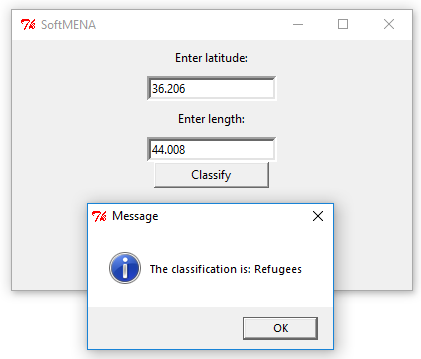}\tabularnewline
\end{tabular}
\par\end{centering}
\caption{Classification of refugees areas in Iraq: a) Duhok, b) Hila, c) Sulem,
d) Saladino, e) Tel Afar, f) Ninaua, g) Kerbala, and h) Erbil.}
\end{figure}

\begin{figure}
\begin{tabular}{cc}
a)\includegraphics[scale=0.5]{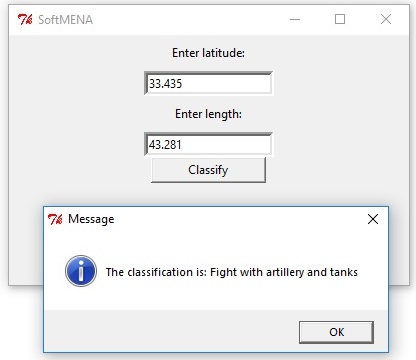} & b)\includegraphics[scale=0.5]{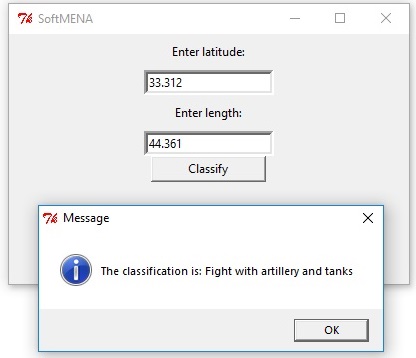}\tabularnewline
c) \includegraphics[scale=0.5]{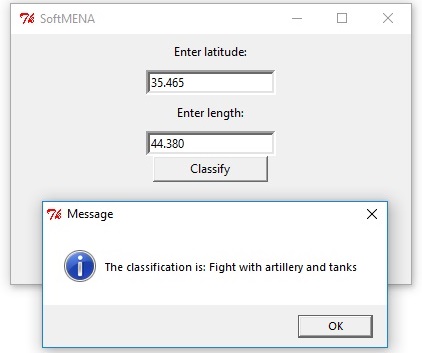} & d) \includegraphics[scale=0.5]{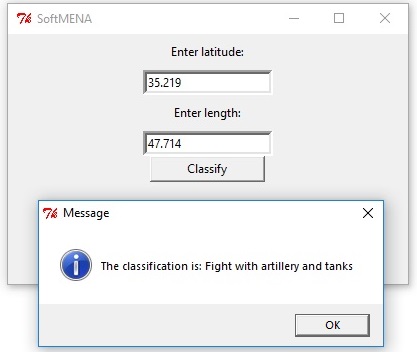}\tabularnewline
e) \includegraphics[scale=0.5]{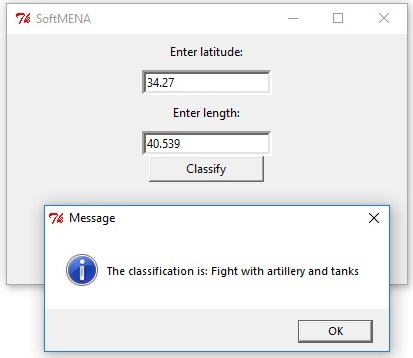} & f) \includegraphics[scale=0.5]{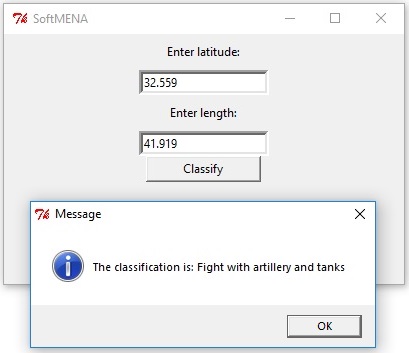}\tabularnewline
g) \includegraphics[scale=0.5]{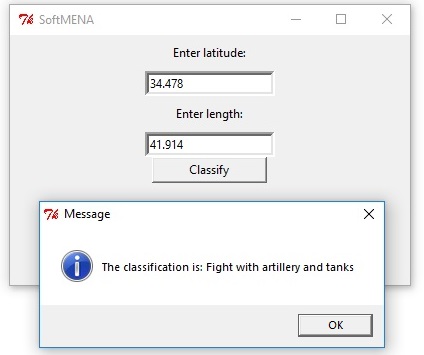} & \tabularnewline
\end{tabular}

\caption{Classification of fights with artillery and tanks areas in Iraq: a)
Ramadi, b) Bagdad, c) Kirkuk, d) Al-Hawija, e) Border Al-Kaim, f)Ambar,
and g) Rawa. }
\end{figure}

\section{Deployment and Feedback}

During the deployment phase, a software was created to classify the
areas of Iraq related to the events obtained with the Decision Trees
algorithm, for refugees and combat events with artillery and tanks.
The software was built with Python 2.7.12 and the user interface was
created with the TKinter library.

Figure 11 presents the use case diagram of this software. The use
cases are described as follows: 1) \emph{run classifier:} the data
scientist creates a classification model with the decision trees algorithm;
2) \emph{evaluate classifier:} the data scientist evaluates the classification
model by means of cross validation; 3) \emph{input longitude and latitude:}
the end user inputs in the graphical user interface the latitude and
longitude of Iraq that he/she wishes to classify; 4) \emph{read the
model for classification:} the algorithm processes the input longitude
and latitude values; 5) \emph{obtain the classification: }the software
obtains the classification of the area. A video shows the software
in action \footnote{https://vimeo.com/268047670}.

\begin{figure}[H]
\begin{centering}
\includegraphics[width=0.5\columnwidth]{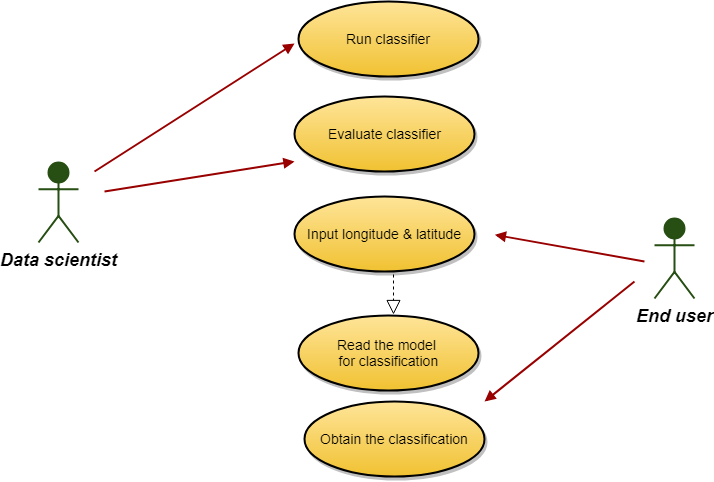}
\par\end{centering}
\caption{Use cases diagram.}
\end{figure}

\section{Conclusions and Future Work}

In this research work we applied data science to discover violence-related
issues in Iraq despite the lack of open governmental data. Specifically
we applied the KNN, Näive Bayes, Decision Trees, and Logistic Regression
classification algorithms to discover violence-related social issues
in Iraq with open data from GDELT in terms of refugees, humanitarian
aid, violent protests, fights with artillery and tanks, and mass killings.
The best results were obtained with the Decision Trees algorithm to
discover areas with refugee crises and artillery fights. A software
prototype was created to show the feasibility of our approach. This
software classifies the zones of Iraq with available or unavailable
data by using the latitude and longitude values of the area to be
studied. The results were compared with official maps. The results
are promising when comparing the results of the classifier with these
maps.

As future work we expect to extend the analysis to other countries
in the Middle East. Also, we are planning to use Apache Spark to process
GDELT data in real time via stream processing. In addition, we are
working on improving the software interface. Specifically, we want
the user interface to show a dynamic map in which the user will be
able to make classifications by clicking on an area of interest on
the map.

\section*{Declaration of Competing Interest}

On behalf of all authors, the corresponding author states that there
is no conflict of interest.

\section*{References}

\bibliographystyle{IEEEtran}
\bibliography{Manuscript.bib}

\end{document}